\documentclass[12pt]{article}
\catcode`\@=11
\@addtoreset{equation}{section}

\global\arraycolsep=2pt
\usepackage{mathrsfs,amsbsy,amssymb,latexsym,amsfonts,amsmath,cite}
\usepackage{graphicx,color}

\newcommand\no{\nonumber}

\newcommand{\dd}{{\rm d}}

\newcommand{\bnabla}{\bar{\nabla}}
\newcommand{\bg}{\bar{g}}
\newcommand{\bphi}{\bar{\phi}}
\newcommand{\bpsi}{\bar{\psi}}
\newcommand{\bR}{\bar{R}}

\allowdisplaybreaks

\begin{document}

\begin{flushright}
\parbox{4cm}
{KUNS-2765 \\
YITP-19-48 
}
\end{flushright}

\vspace*{0.2cm}

\begin{center}
{\Large \bf Gravitational perturbations as $T\bar{T}$-deformations \vspace*{0.1cm}\\ 
in 2D dilaton gravity systems
}
\vspace*{0.5cm}\\
{\large  Takaaki Ishii$^{\sharp}$\footnote{E-mail:~ishiitk@gauge.scphys.kyoto-u.ac.jp}, 
Suguru Okumura$^{\sharp}$\footnote{E-mail:~s.okumura@gauge.scphys.kyoto-u.ac.jp}, \\  
Jun-ichi Sakamoto$^{\flat,\natural}$\footnote{E-mail:~junichi.sakamoto@yukawa.kyoto-u.ac.jp}, 
and Kentaroh Yoshida$^{\sharp}$\footnote{E-mail:~kyoshida@gauge.scphys.kyoto-u.ac.jp}} 
\end{center}

\vspace*{0.4cm}

\begin{center}
$^\sharp${\it Department of Physics, Kyoto University, \\ 
Kitashirakawa Oiwake-cho, Kyoto 606-8502, Japan} \vspace*{0.3cm}\\ 
$^\flat${\it Yukawa Institute for Theoretical Physics, Kyoto University, \\ 
Kitashirakawa Oiwake-cho, Kyoto 606-8502, Japan} \vspace*{0.3cm}\\ 
$^\natural${\it Osaka City University Advanced Mathematical Institute (OCAMI), \\ 
3-3-138, Sugimoto, Sumiyoshi-ku, Osaka, 558-8585, Japan} 

\end{center}

\vspace{0.3cm}

\begin{abstract}
We consider gravitational perturbations of 2D dilaton gravity systems 
and show that these can be recast into $T\bar{T}$-deformations 
(at least) under certain conditions, where $T$ means the energy-momentum tensor 
of the matter field coupled to a dilaton gravity. In particular, the class of theories 
under this condition includes a Jackiw-Teitelboim (JT) theory 
with a negative cosmological constant including conformal matter fields. 
This is a generalization of the preceding work on the flat-space JT gravity 
by S.~Dubovsky, V.~Gorbenko and M.~Mirbabayi [arXiv:1706.06604]. 

\end{abstract}

\setcounter{footnote}{0}
\setcounter{page}{0}
\thispagestyle{empty}

\newpage

\tableofcontents

\section{Introduction}

A significant subject is to study integrable deformations of 2D integrable 
quantum field theories (IQFTs) like sine-Gordon models and $O(N)$ vector models\footnote{In the following, we will assume the Lorentz symmetry and consider relativistic unitary theories only.}. An example that has been 
investigated vigorously in recent years is specified 
by the energy-momentum tensor $T$ and often called the $T\bar{T}$-deformation \cite{SZ,Tateo}. 
This deformation is actually given by a determinant of $T$\,, not $T\bar{T}$\,, but 
it is conventionally called the $T\bar{T}$-deformation. The origin of the name would be that the deformation of conformal field theory (CFT)  is eventually described by $T\bar{T}$\,. 
A peculiar feature of $T\bar{T}$-operator has been originally realized in \cite{Z}. 
Then it has been presented in closed form and studied systematically \cite{Bonelli}. 
For a concise review, see \cite{review}. 

\medskip 

In 2D IQFTs, any $N$-body S-matrix is factorized into 
a product of 2-body S-matrices. This factorization property is the onset of  
the quantum integrability. The 2-body S-matrix is a representation of quantum $R$-matrix 
satisfying the Yang-Baxter equation and is determined by Lorentz symmetry, crossing symmetry, 
and unitarity, up to the Castillejo-Dalitz-Dyson (CDD) factor \cite{CDD}. The CDD factor has to be determined 
case by case, depending on the models we are concerned with. This is the usual S-matrix 
bootstrap program in 2D IQFTs. In this context, the $T\bar{T}$-deformations are irrelevant 
deformations and then modify only the CDD factor.  
Hence this factorization property is preserved, and thus the $T\bar{T}$-deformations 
may be called integrable deformations in the usual sense. 

\medskip 

An amazing observation is that one may consider the $T\bar{T}$-deformation of general QFTs 
apart from the integrability. This was pointed out in \cite{Z} and further elaborated in \cite{SZ,Tateo}. 
The deformation effect appears only in the modification of the CDD factor in the case of 
2D IQFTs. Then it would be reasonable to anticipate that the modification due to 
the $T\bar{T}$-deformation appears only as some multiplicative factor to the S-matrix 
even for the general (non-integrable) QFT cases, where the S-matrix factorization 
does not occur any more. One would be able to see works by S.~Dubovsky et al. 
\cite{Dubovsky1,Dubovsky2,Dubovsky} as a support for this anticipation. 

\medskip 

One may find out an interesting application of $T\bar{T}$-deformation in the context of 2D dilaton 
gravity called the Jackiw-Teitelboim (JT) gravity \cite{Jackiw,Teitelboim} 
(For comprehensive reviews, see \cite{V,Nojiri}). 
In the pioneering work \cite{Dubovsky}, a gravitational perturbation around a solution in the flat-space 
JT gravity has been considered. The perturbation can be reinterpreted as a $T\bar{T}$-deformation of the original matter 
action. Then the matter theory gets a gravitationally dressing factor in front of the 
S-matrix due to the perturbation. This result indicates that the (classical) gravitational perturbation 
can be seen as a non-perturbative quantum effect to the matter sector, and 
the deformation effect can be computed rigorously while the S-matrix of the original theory cannot 
be evaluated exactly in general. However, this intriguing result has been shown only 
in the flat-space JT gravity with a simple dilaton potential. It should be significant 
to figure out to what extent this result should be valid.   

\medskip

In this paper, we study gravitational perturbations of 2D dilaton gravity systems with matter fields 
in a more general setup. Then we show that these perturbations can be seen 
as $T\bar{T}$-deformations (at least) under certain conditions, 
where $T$ means the energy-momentum tensor 
of the matter field coupled to the dilaton gravity. 
In particular, the class of theories under this condition 
includes a JT gravity with a negative cosmological constant with conformal matter fields. 
This is a generalization of the work \cite{Dubovsky} and has potential applications in the context of 
the AdS$_2$ holography \cite{AP,Jensen,MSY,EMV}.

\medskip 

This paper is organized as follows. In section 2, we introduce 2D dilaton-gravity systems coupled 
with an arbitrary matter field. In section 3, gravitational perturbations in the systems 
are considered and the equations of motion for the fluctuations are derived. 
In section 4, (at least) for some cases, it is shown that the gravitational perturbations 
can be regarded as $T\bar{T}$-deformations of the original matter Lagrangian.  
Section 5 is devoted to conclusion and discussion. In Appendix A, we list some useful formulae 
to compute gravitational perturbations in a covariant way. Appendix B explains how to derive 
the quadratic action in terms of the fluctuation in detail.

\section{2D dilaton gravity systems coupled with matter field}

In the following, we will consider a 2D dilaton gravity system coupled with an arbitrary matter 
field $\psi$\,. We will work with the Lorentzian signature, and the coordinates are described 
as $x^{\mu}=(x^0,x^1)=(t,x)$\,. The metric field and dilaton are given 
by $g_{\mu\nu}(x^{\mu})$ and $\phi(x^{\mu})$\,, respectively. 

\medskip
 
The classical action is given by 
\begin{align}
S[g_{\mu\nu},\phi,\psi]&=S_{\rm dg}[g_{\mu\nu},\phi] + S_{m}[\psi,g_{\mu\nu},\phi]\,, \label{action}\\ 
\begin{split}
S_{\rm dg}[g_{\mu\nu},\phi] 
&=\frac{1}{16\pi G_N}\int\! \dd^2 x\,\sqrt{-g}\,\left[\,\phi\, R-U(\phi)\,\right]\,,
\\
S_{m}[\psi,g_{\mu\nu},\phi] &=\int\! \dd^2 x\,\sqrt{-g}\,F(\phi)\,\mathcal{L}_{m}[g_{\mu\nu}, 
\psi]\,,\label{eq:action}
\end{split}
\end{align}
where $G_{N}$ is the Newton constant in two dimensions and $U(\phi)$ is a dilaton potential. 
The matter action $S_{m}$ may include a non-trivial dilaton coupling $F(\phi)$ in general in front of the 
matter Lagrangian $\mathcal{L}_m$\,. In the following, we assume that $F(\phi)$ is constant 
and normalized as $F(\phi)=1$\,, for simplicity. 

\medskip

The equations of motion of this system are given by 
\begin{align}
&R-U'(\phi)  =0\,,\label{eq:eom1}\\
&\frac{1}{2}g_{\mu\nu}U(\phi)
-\left(\nabla_{\mu}\nabla_{\nu}\phi-g_{\mu\nu}\nabla^2 \phi\right)
=8\pi G_{N}\,T_{\mu\nu}\,,\label{eq:eom2}
\end{align}
where we have defined the energy-momentum tensor $T_{\mu\nu}$ for the matter field $\psi$ as 
\begin{align}
T_{\mu\nu} \equiv -\frac{2}{\sqrt{-g}}\frac{\delta S_m}{\delta g^{\mu\nu}}\,.
\end{align}
We do not discuss the dynamics of the matter field itself (nor the backreaction of the metric 
and dilaton to the matter field). Therefore the equation of motion for $\psi$ is not included.
In deriving the equations of motion (\ref{eq:eom2}) for the metric,
we have used the fact that the Einstein tensor $G_{\mu\nu}$ in two dimensions vanishes: 
\begin{align}
G_{\mu\nu} \equiv R_{\mu\nu}-\frac{1}{2}g_{\mu\nu} R=0\,. \label{eq:2dEin}
\end{align}

\medskip

In our later discussion, we are interested in studying gravitational perturbations around 
a vacuum solution (i.e., a solution obtained when $T_{\mu\nu}=0$)\footnote{Note here 
that the dilaton is not regarded as a matter field but a part of the metric. 
This viewpoint would be rather natural as some dilaton gravities are obtained 
by dimensional reduction of higher-dimensional theories.}. 
Hence, it would be useful to write down some relations 
for the dilaton $\phi$ in an arbitrary vacuum solution. 
When $T_{\mu\nu}=0$\,,
the equation of motion (\ref{eq:eom2}) for the metric takes a simple form
\begin{align}
\nabla_{\mu}\nabla_{\nu}\phi=g_{\mu\nu}\left(\nabla^2 \phi+\frac{1}{2}U(\phi)\right)
\,.\label{eq:eom1-vacuum}
\end{align}
The trace of (\ref{eq:eom1-vacuum}) is given by 
\begin{align}
&\nabla^2 \phi+U(\phi)=0\,.\label{eq:eom-phitrace}
\end{align}
By using (\ref{eq:eom-phitrace}) and (\ref{eq:eom1-vacuum}), 
the dilaton potential $U(\phi)$ can be deleted. The resulting expression is  
\begin{align}
\nabla_{\mu}\nabla_{\nu}\phi=\frac{1}{2}g_{\mu\nu} \nabla^2 \phi\,.  
\label{eq:eom-phimunu}
\end{align}

\subsection*{Comment on the flat-space JT gravity}  

In the work \cite{Dubovsky}, S.~Dubovsky et al considered a special case 
called the flat-space JT gravity. This case corresponds to the following dilaton potential
\begin{align}
U(\phi)=\Lambda\,,\label{eq:dil-potential-const}
\end{align}
where $\Lambda$ is a constant.
The vacuum solution is uniquely determined (up to trivial ambiguities) as 
\begin{align}
\dd^2 s=  -2\dd x^+\dd x^-\,,\qquad 
\phi = \frac{\Lambda}{2}x^+  x^-\,, 
\label{eq:flat-bg}
\end{align}
where the light-cone coordinates are defined as 
\[
x^{\pm} \equiv \frac{1}{\sqrt{2}}(t \pm x)\,. 
\]
In \cite{Dubovsky}, the dilaton gravity system coupled with an arbitrary matter field has been expanded 
around this vacuum solution and the quadratic fluctuations have been recast into 
a form of $T\bar{T}$-deformation. We will return to this point as a special example 
later after we carry out general computation.

\section{Perturbing 2D dilaton gravity systems}

By starting from the classical action (\ref{action}), let us consider a gravitational perturbation 
around a vacuum solution. In the following, we will slightly change our notation. The original metric, 
dilaton and matter field are described as $g_{\mu\nu}$\,, $\phi$ and $\psi$\,, respectively.  The vacuum solution, 
which is taken as an expansion point, is specified by $\bg_{\mu\nu}$ and $\bphi$\,. Since we assume that the expansion point is a vacuum solution with $T_{\mu\nu}=0$\,, the matter field $\psi$ should be regarded as a fluctuation (i.e., $\psi$ should be expanded around zero). 
In summary, a gravitational perturbation around a vacuum solution is described as  
\begin{align}
g_{\mu\nu}=\bg_{\mu\nu}+h_{\mu\nu}\,, 
\qquad\phi=\bphi+\sigma\,,\qquad \psi = 0 + \psi\,, 
\label{eq:fluctuation}
\end{align}
where $h_{\mu\nu}$ and $\sigma$ are fluctuations of metric and dilaton, respectively, and $\psi$ in the right hand side is treated as a fluctuation with a slight abuse of notations. 
Note here that since $\bg_{\mu\nu}$ and $\bphi$ 
describe a vacuum solution, the equations of motion (\ref{eq:eom1-vacuum}) should be satisfied. 

\medskip 

It is an easy practice to derive a vacuum solution explicitly by specifying a dilaton potential at the beginning. 
However, we will not do that here and keep an abstract form of the vacuum solution so as to argue in a {\it covariant} way. If we use a concrete expression of the vacuum solution, 
covariance of the expression is not manifest like in \cite{Dubovsky}.

\subsection{The quadratic action}

Let us expand the classical action $S[g_{\mu\nu}, \phi, \psi]$ in (\ref{action}) 
by the fluctuations (\ref{eq:fluctuation}).
The classical action can be expanded as
\begin{align}
S[g_{\mu\nu}, \phi, \psi] &= S^{(0)}+S^{(1)}+S^{(2)}+\cdots\, \no \\ 
&= S^{(0)}_{\rm dg}[\bg_{\mu\nu},\bphi] + S^{(1)}_{\rm dg}[\bg_{\mu\nu},\bphi;h_{\mu\nu},\sigma] 
+ S^{(2)}_{\rm dg}[\bg_{\mu\nu},\bphi;h_{\mu\nu},\sigma]  \no \\ 
& ~~~+ S^{(1)}_{\rm m}[\bg_{\mu\nu};\psi] 
+ S^{(2)}_{\rm m}[\bg_{\mu\nu};\psi,h_{\mu\nu}] + \cdots \,,
\label{eq:action-expanded}
\end{align}
where the superscript of $S^{(n)}$ denotes the order of fluctuations. The zeroth order part 
$S^{(0)}_{\rm dg}$ is the classical value of $S_{\rm dg}$ with the vacuum configuration. 
It is just a constant in the case of \cite{Dubovsky} but in general depends on the coordinates as we will see later.
Then the first order action $S^{(1)}_{\rm dg}$ should vanish since the vacuum solution satisfies 
the equations of motion with $\bpsi =0$\,. For the matter sector, $S_{\rm m}^{(1)}$ describes the matter field action 
on the classical background specified by the metric of the vacuum solution. 
The second-order contribution $S_{\rm m}^{(2)}$ is evaluated as 
\begin{equation}
S_{\rm m}^{(2)} = \delta g^{\mu\nu} 
\left.\frac{\delta S_{\rm m}}{\delta g^{\mu\nu}}
\right|_{g_{\mu\nu} =\bg_{\mu\nu}}
 = \frac{1}{2}\int\!\dd^2 x\,\sqrt{-\bg}\, h^{\mu\nu} t_{\mu\nu}\,, 
\end{equation}
where $t_{\mu\nu}$ is the energy-momentum tensor for the matter theory described 
by $S_{\rm m}^{(1)}$\,. Note here that 
\[
g^{\mu\nu} =\bg^{\mu\nu} - h^{\mu\nu} + \mathcal{O}(h^2)\,. 
\]
where the indices in the perturbations are raised, lowered, and contracted with the background metric $\bg_{\mu\nu}$: $h^{\mu\nu} \equiv \bg^{\mu\rho}\bg^{\nu\sigma}h_{\rho\sigma}$\,. 

\medskip

After carrying out a lengthy calculation, we obtain the explicit expression of the quadratic action 
$S^{(2)} \equiv S^{(2)}_{\rm dg} + S^{(2)}_{\rm m}$\,.
By ignoring total derivative terms, this is given by
\begin{align}
S^{(2)}&=\frac{1}{16\pi G_N}\int\! \dd^2 x\,
\sqrt{-\bg}\biggl(\left[\bnabla^{\mu}\bnabla^{\nu}h_{\mu\nu}-\bnabla^2 h
-\frac{1}{2}h\,U'(\bphi)
-\frac{1}{2}U''(\bphi) \sigma\right]\sigma
\no\\
&\quad
-\frac{1}{8}\bnabla^{2}\bphi\,h_{\mu\nu}h^{\mu\nu}
-\bnabla^{\rho}\bphi \left[
-\frac{1}{2}h_{\rho\sigma} \bnabla_{\mu}h^{\mu\sigma}
+\frac{1}{4}h \bnabla^{\mu}h_{\mu\rho}
+\frac{3}{4}h_{\rho\mu}\bnabla^{\mu}h
\right]\biggr)\no\\
&\quad+\frac{1}{2}\int\! \dd^2 x\,\sqrt{-\bg}\,h^{\mu\nu} t_{\mu\nu}\,.
\label{eq:quadratic-action}
\end{align}
where $h \equiv \bg^{\mu\nu}h_{\mu\nu}$.
The derivation of the above expression is given in Appendix \ref{sec:der-qaction}.

\subsection{Equations of motion for the fluctuations}

Taking the variation of the quadratic action \eqref{eq:action-second-1}, or expanding 
the equations of motion \eqref{eq:eom1} and \eqref{eq:eom2}, 
we obtain the equations of motion for the fluctuations as
\begin{align}
&\bnabla^{\mu}\bnabla^{\nu}h_{\mu\nu}-\bnabla^2 h
-\frac{1}{2}U'(\bphi)\,h -U''(\bphi) \sigma=0\,,\label{eq:eom-sigma}\\
&\left(-\bnabla_{\mu}\bnabla_{\nu}\sigma+\bg_{\mu\nu}\bnabla^2 \sigma
+\frac{1}{2}\bg_{\mu\nu}\,
U'(\bphi)\,\sigma\right)
+\frac{1}{2}\bnabla^2\phi\,(h_{\mu\nu}-\bg_{\mu\nu}\,h)\no\\
&+\frac{1}{2}\bnabla^\rho \bphi\left[\left(\bnabla_{\mu}h_{\rho \nu}+\bnabla_{\nu} h_{\rho \mu}
-\bnabla_{\rho}h_{\mu\nu}\right)
-2\bg_{\mu\nu}\left(\bnabla^{\sigma}h_{\rho\sigma}-\frac{1}{2}\bnabla_{\rho}h\right)\right]
=8\pi G_{N}\,t_{\mu\nu}\,.
\label{eq:eom-h2}
\end{align}
Taking the trace of (\ref{eq:eom-h2}) leads to 
\begin{align}
&\left(\bnabla^2 +U'(\bphi)\right)\sigma+\frac{1}{2}U(\bphi)\,h
-\bnabla^\rho \bphi\left(\bnabla^{\sigma}h_{\rho\sigma }
-\frac{1}{2}\bnabla_{\rho}h\right)
=8\pi G_{N}\,t^{\mu}_{\mu}\,.
\label{eq:eom-h-trace}
\end{align}
Subtracting the trace part (\ref{eq:eom-h-trace}) from (\ref{eq:eom-h2}), we obtain
\begin{align}
&\bnabla_{\mu}\bnabla_{\nu}\sigma+\frac{1}{2}\bg_{\mu\nu}\,U'(\bphi)\,\sigma\no\\
&=-8\pi G_{N}\,(t_{\mu\nu}-\bg_{\mu\nu}t^{\rho}_{\rho})-\frac{1}{2}U(\bphi)\,h_{\mu\nu}
+\frac{1}{2}\bnabla^\rho \bphi\left(\bnabla_{\mu}h_{\rho \nu}+\bnabla_{\nu} h_{\rho \mu}
-\bnabla_{\rho}h_{\mu\nu}\right)\,,\label{eq:eom-usefull}
\end{align}
which may also be useful in our discussion later. 

\medskip

Finally, it would be worth noting why we employed \eqref{eq:action-second-1} rather than  
(\ref{eq:quadratic-action}).  
As already pointed out, the Einstein tensor vanishes in two dimensions. Hence there may be some non-
trivial relations among fluctuations. If such relations are utilized before taking variations with the 
fluctuations, then the forms of 
equations of motion become different. Eventually, if we derive the equations of motion from 
(\ref{eq:quadratic-action}), then the resulting expression is different from the ones obtained by expanding 
the equations of motion \eqref{eq:eom1} and \eqref{eq:eom2}, though these are equivalent under the 
relations intrinsic to two dimensions. However, if we start from \eqref{eq:action-second-1}, 
the result is the same as the one obtained from \eqref{eq:eom1} and \eqref{eq:eom2},  
without using any special relations.

\section{Gravitational perturbations as $T\bar{T}$-deformations}

In this section, we will explore whether or not the quadratic action can be regarded as a $T\bar{T}$ deformation of the original matter action in more general cases. At least under some conditions, the gravitational perturbations 
can be seen as $T\bar{T}$-deformations as explicitly shown below.

\subsection{The case of the flat-space JT gravity}

As the first example, let us revisit the case of the flat-space JT gravity considered in \cite{Dubovsky}. 
This case is realized by taking a constant dilaton potential
\begin{align}
U'(\phi)=0\,,\qquad U(\phi)=\Lambda\,,  
\end{align}
where $\Lambda$ is a constant. 
The background dilaton $\bphi$ should satisfy the following conditions 
\begin{align}
\bR=0\,,\qquad 
\bnabla^2\bphi=-U(\bphi)=-\Lambda\,, 
\end{align}
which follow from (\ref{eq:eom1}) and (\ref{eq:eom1-vacuum}).

\medskip

The first is to solve the equation of motion (\ref{eq:eom-sigma})\,, which in the present case 
is simplified to 
\begin{align}
&\partial^{\nu}(\partial^{\mu}h_{\mu\nu}-\partial_{\nu} h)=0\,.\label{eq:eom-sigma-Dub}
\end{align}
A possible solution to the equation (\ref{eq:eom-sigma-Dub}) is given by 
\begin{align}
h_{\mu\nu}=-16\pi G_{N}(t_{\mu\nu}-\bg_{\mu\nu}t^{\rho}_{\rho})\, k \,,
\label{eq:metric-ansatz}
\end{align}
where $k$ is an overall constant. It is easy to see that the above $h_{\mu\nu}$ indeed solves 
the equation (\ref{eq:eom-sigma-Dub}) by noting that the conservation law of 
the energy-momentum tensor $t_{\mu\nu}$, $\bnabla_\mu t^{\mu\nu}=0$, leads to the relation
\begin{align}
\partial^{\mu} h_{\mu\nu}=\partial_{\nu} h\,.  
\label{flat}
\end{align}

\medskip

The next is to construct an explicit solution of the dilaton fluctuation $\sigma$ 
under the metric solution (\ref{eq:metric-ansatz}).
By using the conservation law of  the energy-momentum tensor,
the equations of motion (\ref{eq:eom-usefull}) can be rewritten as 
\begin{align}
\partial_{\mu}\partial_{\nu}\sigma&=-8\pi G_{N}\,(t_{\mu\nu}-\bg_{\mu\nu}t^{\rho}_{\rho})
-\frac{1}{2}U(\bphi)\,h_{\mu\nu}
+\frac{1}{2}\partial^\rho \bphi\left(\partial_{\mu}h_{\rho \nu}+\partial_{\nu} h_{\rho \mu}
-\partial_{\rho}h_{\mu\nu}\right)\no\\
& =-8\pi G_{N}
\biggl[\left(1-k \Lambda\right)-\frac{k \Lambda}{2} x^{\rho} \partial_{\rho} \biggr]
\left(t_{\mu\nu}-\bg_{\mu\nu}\,t^{\sigma}_{\sigma}\right)\,.
\label{eq:ddsigmaD}
\end{align}
It is possible to construct explicitly a non-local solution to the equations (\ref{eq:ddsigmaD}).
To see this, let us first decompose the dilaton into two parts as 
\begin{align}
\sigma(x^+,x^-)&=\sigma_{0}(x^+,x^-)+\sigma_{\text{non-local}}(x^+,x^-)\,. 
\label{eq:sigma-0nonlocal}
\end{align}
Here the first term $\sigma_{0}(x^+,x^-)$ corresponds to the sourceless part,
\begin{align}
\sigma_{0}(x^+,x^-)=a_1+a_2\,x^++a_3\,x^-\,,\qquad a_{i}~(i=1,2,3) \mbox{:~arbitrary real consts.}, 
\end{align}
and obviously satisfies $\partial_{\mu}\partial_{\nu}\sigma=0$\,.
The second term $\sigma_{\text{non-local}}(x^+,x^-)$ describes 
the non-local part,
\begin{align}
\sigma_{\text{non-local}}&=
4\pi G_{N}\Biggl[k\Lambda\int^{x^+}_{0}\!\!\!\dd s\,s\,t_{++}(s,x^-)+k\Lambda
\int^{x^-}_{0}\!\!\!\dd s\,s\,t_{--}(x^+,s)\no\\
&\quad -2\left(k\Lambda-1\right)\int^{x^+}_{0}\!\!\dd s\int^{x^-}_{0}\!\!\dd s'\,t_{+-}(s,s')\no\\
&\quad +\left(k\Lambda-2\right)
\left(\int^{x^+}_{u^+_1}\!\!\dd s\int^{s}_{u^+_2}\!\!\dd s'\,t_{++}(s',0) 
+\int^{x^-}_{u^-_1}\!\!\dd s\int^{s}_{u^{-}_2}\!\!\dd s'\,t_{--}(0,s')\right)
\Biggr]\,,\label{eq:sigma-non-local-d}
\end{align}
where $u^{\pm}_{1,2}$ are arbitrary constants.\footnote{%
The domain of integration may change due to the shift symmetry of the background, 
$x^+ \to x^+ - a^+$ and $x^- \to x^- - a^-$\,. After making this shift, the background dilaton 
is transformed like 
\begin{equation}
\bar{\phi} = \frac{\Lambda}{2} (x^+ - a^+) (x^- - a^-)\,.
\end{equation}
Then the non-local part \eqref{eq:sigma-non-local-d} is modified as
\begin{align}
\sigma_{\text{non-local}}&=
4\pi G_{N}\Biggl[k\Lambda\int^{x^+}_{a^+}\!\!\!\dd s\,(s-a^+)\,t_{++}(s,x^-)
+k\Lambda\int^{x^-}_{a^-}\!\!\!\dd s\,(s-a^-)\,t_{--}(x^+,s)\no\\
&\quad -2\left(k\Lambda-1\right)\int^{x^+}_{a^+}\!\!\dd s\int^{x^-}_{a^-}\!\!\dd s'\,t_{+-}(s,s')\no\\
&\quad +\left(k\Lambda-2\right)
\left(\int^{x^+}_{u^+_1}\!\!\dd s\int^{s}_{u^+_2}\!\!\dd s'\,t_{++}(s',a^-) 
+\int^{x^-}_{u^-_1}\!\!\dd s\int^{s}_{u^{-}_2}\!\!\dd s'\,t_{--}(a^+,s')\right)
\Biggr]\,.
\label{eq:sigma-non-local-d-shifted}
\end{align}
}
It is easy to check that the non-local solution (\ref{eq:sigma-0nonlocal}) satisfies the 
equations of motion  
\eqref{eq:ddsigmaD} by using the conservation law of the energy-momentum tensor. 
Note here that the sign of the deformation depends on the values of $\Lambda$ and $k$\,. 

\medskip

After substituting the solutions (\ref{eq:metric-ansatz}) and  (\ref{eq:sigma-0nonlocal})   
into (\ref{eq:quadratic-action}), the resulting quadratic action is given by 
(up to the total derivative terms)
\begin{align}
S^{(2)}&=\frac{1}{16\pi G_N}\int\! \dd^2 x\,
\biggl(-\frac{1}{8}\partial^{2}\bphi\,h_{\mu\nu}h^{\mu\nu}
-\frac{1}{4}\partial^{\rho}\bphi \left(
h \partial_{\rho}h
+h_{\rho\mu}\partial^{\mu}h
\right)+8\pi G_N\,h^{\mu\nu}t_{\mu\nu}\biggr)\no\\
&=\frac{1}{16\pi G_N}\int\! \dd^2 x\,
\biggl(-\frac{1}{8}\partial^{2}\bphi\,h_{\mu\nu}h^{\mu\nu}
+\frac{1}{4}\partial^{\mu}\partial^{\rho}\bphi\,
h_{\rho\mu}h
+8\pi G_N\,h^{\mu\nu}t_{\mu\nu}
\biggr)\no\\
&=-16\pi G_N\left(\frac{1}{2k}-\frac{\Lambda}{8}\right)\, k^2
\int\! \dd^2 x \left[\,t_{\mu\nu}t^{\mu\nu}-(t^{\mu}_{\mu})^2\,\right]\,.
\label{eq:qS-ttbar}
\end{align}
Thus the quadratic action can be regarded as a $T\bar{T}$ deformation of $S^{(1)}_{\rm m}$\,. 
It should be remarked here that the quadratic action (\ref{eq:qS-ttbar}) can be derived by using 
only the expression of $h_{\mu\nu}$\,, without using the explicit expression of $\sigma$\,. 
It is because the first line proportional to $\sigma$ in the action (\ref{eq:quadratic-action}) 
vanishes identically under the condition (\ref{flat}) and then the action is independent of $\sigma$\,, 
though the existence of $\sigma$ as a consistent solution to (\ref{eq:ddsigmaD})
is crucial as carefully discussed in \cite{Dubovsky}. 

\medskip 

If we set $k=\frac{2}{\Lambda}$ as in \cite{Dubovsky}, then 
the resulting action is simplified to 
\begin{align}
S^{(2)}
=-\frac{8\pi G_{N}}{\Lambda}
\int\! \dd^2 x \left[\,t_{\mu\nu}t^{\mu\nu}-(t^{\mu}_{\mu})^2\,\right]\,.
\end{align}
This is nothing but the result obtained in \cite{Dubovsky}.

\subsection{The $U'(\phi) \neq 0$ and $U''(\phi) = 0$ case}

The next case is a more general class of 2D dilaton gravity systems with dilaton potentials 
satisfying the following conditions: 
\begin{align}
U'(\phi)\neq 0\,,\qquad U''(\phi)=0\,. 
\label{eq:TT-condition}
\end{align}
Under these conditions, the equations of motion for the fluctuations are simplified to 
\begin{align}
&\bnabla^{\mu}\bnabla^{\nu}h_{\mu\nu}-\bnabla^2 h-\frac{1}{2}U'(\bphi)\,h=0\,, 
\label{eq:eom-flucdil-ap}\\
&\left(-\bnabla_{\mu}\bnabla_{\nu}\sigma+\bg_{\mu\nu} \bnabla^2 \sigma 
+\frac{1}{2}\bg_{\mu\nu}\,U'(\bphi)\,\sigma\right)
+\frac{1}{2}\bnabla^2\bphi\,(h_{\mu\nu}-\bg_{\mu\nu}\,h)
\label{eq:eom-flucmet-ap}
\no\\
&+\frac{1}{2}\bnabla^\rho \bphi\left[\left(\bnabla_{\mu}h_{\rho \nu}
+\bnabla_{\nu} h_{\rho \mu}-\bnabla_{\rho}h_{\mu\nu}\right)
-2\bg_{\mu\nu}\left(\bnabla^{\sigma}h_{\rho\sigma}-\frac{1}{2}\bnabla_{\rho}h\right)\right]
=8\pi G_{N}\,t_{\mu\nu}\,.
\end{align}

\medskip

As in the previous case, let us solve the first equation (\ref{eq:eom-flucdil-ap}). 
Suppose that the matter $\psi$ is taken to be a conformal matter i.e. $t^{\mu}_{\mu}=0$\,. 
Then it is easy to find out a solution to the equation (\ref{eq:eom-flucdil-ap}), 
\begin{align}
h_{\mu\nu}=-k\cdot 16\pi G_{N}\,t_{\mu\nu}\,,\qquad h=-k \cdot 16\pi G_{N}\,t^{\mu}_{\mu}=0\,, 
\end{align}
with the help of the conservation law of the energy-momentum tensor. 
Finally, the quadratic action (\ref{eq:quadratic-action}) can be rewritten as 
\begin{align}
S^{(2)}
&=(16\pi G_N)\,\frac{k^2}{8}\int\! \dd^2 x\,
\sqrt{-\bg}\,\left(U(\bphi)-\frac{4}{k}\right)\,t_{\mu\nu}t^{\mu\nu}
\,.
\end{align}
Thus this can also be regarded as a $T\bar{T}$ deformation.
Note that the coefficient of $t_{\mu\nu}t^{\mu\nu}$ is linear in terms of the background dilaton $\bphi$\,.

\subsection*{Concrete examples }

In the following, we show two examples for the present case. 

\subsubsection*{(i) the Almheiri-Polchinski model}

An interesting example of $2$D dilaton gravity systems satisfying the condition 
(\ref{eq:TT-condition}) is the Almheiri-Polchinski (AP) model \cite{AP}\footnote{Here, 
we dare to call the JT gravity with a conformal matter field as the AP model 
so as to respect the analysis on the conformal matter in \cite{AP} 
which plays a crucial role in our analysis here.}.
This model has the dilaton potential\footnote{As for the notation, 
note that our $\bphi$ corresponds to 
$\Phi^2$ in \cite{AP}.}
\begin{align}
U(\phi)=\Lambda-\frac{2}{L^2}\phi\,.
\end{align}

\medskip 

Let us construct the explicit form of $\sigma$.
In the following, we will employ conformal gauge and the metric is given by 
\begin{align}
\dd^2 s=\bg_{\mu\nu}\dd x^{\mu} \dd x^{\nu}=-2 {\rm e}^{2\bar{\omega}}\dd x^{+} \dd x^{-}\,.
\end{align}
The general vacuum solution incorporates the AdS$_2$ metric and a non-constant dilaton
\begin{align}
{\rm e}^{2\bar{\omega}}
= \frac{2\,L^2}{(x^+-x^-)^2}\,,\qquad \bphi=\frac{L^2}{2}
\left(\Lambda+\frac{a+b\,(x^+ +x^-)+c\,x^+ x^-}{x^+-x^-}\right)\,, 
\end{align}
where $L$ is the AdS radius and $a$, $b$ and $c$ is an arbitrary constant.

\medskip 

Let us solve the equations of motion  (\ref{eq:eom-usefull}). 
The $(++)$ and $(--)$ components of (\ref{eq:eom-usefull}) are evaluated as 
\begin{align}
{\rm e}^{2\bar{\omega}}\partial_+\left({\rm e}^{-2\bar{\omega}}\partial_+\sigma\right)=
& -8\pi G_N {\cal T}_+(x^+)\,,
\label{eq:AP++}\\
{\rm e}^{2\bar{\omega}}\partial_-\left({\rm e}^{-2\bar{\omega}}\partial_-\sigma\right)=&
-8\pi G_N {\cal T}_-(x^-)\,, 
\label{eq:AP--}
\end{align}
where ${\cal T}_{\pm} (x^{\pm})$ are defined as 
\begin{equation}
 {\cal T}_{\pm} (x^{\pm}) \equiv (1 \mp b\,k \mp c\,k\,x^{\pm})\,t_{\pm \pm} 
 \mp \frac{k}{4}\left(a+2b\,x^{\pm} + c\,(x^{\pm})^2\right)\partial_{\pm} t_{\pm \pm}\,. 
 \end{equation}
By following \cite{AP}, it is useful to express $\sigma$ 
with a scalar function $M(x^+\,,x^-)$ as 
\begin{align}
\sigma(x^+,x^-) \equiv \frac{M(x^+\,,x^-)}{x^+-x^-}\,.
\end{align}
Then the left-hand sides of (\ref{eq:AP++}) and (\ref{eq:AP--}) can be written as
\begin{align}
{\rm e}^{2\bar{\omega}}\partial_\pm\left({\rm e}^{-2\bar{\omega}}\partial_\pm\sigma\right)=\frac{\partial_\pm\partial_\pm M(x^+\,,x^-)}{x^+-x^-}\,.
\end{align}
By integrating (\ref{eq:AP++}) and (\ref{eq:AP--}), the general solution can be derived as 
\begin{align}
M(x^+\,,x^-)=I_0(x^+,x^-) + I^+(x^+,x^-) - I^-(x^+,x^-)\,.\label{APsolution}
\end{align}
Here $I_0$ is the sourceless solution, 
\[
I_0(x^+,x^-)\equiv A + B\,(x^++x^-) + C\, x^+x^-\,,  \qquad A,~B,~C \mbox{:~arbitrary real consts.}, 
\] 
and $I^\pm(x^+,x^-)$ correspond to the non-local parts of dilaton and are given by 
\begin{align}
I^\pm(x^+,x^-) \equiv &\, 8\pi G_N 
\int^{x^\pm}_{u^\pm}\!\!\dd s\, (s-x^+)(s-x^-)\,{\cal T}_\pm(s)\,.
\end{align}
The $(+-)$ component of (\ref{eq:eom-usefull}) is drastically simplified 
due to the traceless condition $t_{+-}=0$ and is given by 
\begin{align}
\partial_+\partial_-\sigma+\frac{2\,\sigma}{(x^+-x^-)^2}=&\,0\,.\label{eq:AP+-}
\end{align}
It is an easy task to see that $\sigma$ with (\ref{APsolution}) satisfies 
the above condition (\ref{eq:AP+-}).

\medskip 

It should be remarked that this {\it non-local} solution to (\ref{eq:AP+-}) might be epochal. 
One would usually try to employ hypergeometric functions or Gegenbauer polynomials 
to solve it by assuming that the solution is local. But this solution is non-local and it has not been 
presented at least as far as we know. 
This non-local solution may play an important role in resolving 
the long-standing issue of the AdS$_2$/CFT$_1$ correspondence. 

\subsubsection*{A flat-space limit}

It is intriguing to consider a flat-space limit of the vacuum solution in the AP model  
(See also Appendix B of \cite{Dubovsky} for the limit with the embedding coordinates). 

\medskip 

Let us first take a constant shift of $x^\pm$ and introduce 
new coordinates $X^\pm$ defined as 
\begin{align}
X^\pm \equiv x^\pm \mp\frac{L}{\sqrt{2}}\,.
\end{align}
By taking the large radius limit $L\rightarrow\infty$, the AdS$_2$ metric goes 
to the Minkowski metric, 
\begin{align}
{\rm d}s^2
= \frac{-4\,L^2\dd x^+\dd x^-}{(x^+-x^-)^2} \quad 
\longrightarrow \quad 
-2\,\dd X^+\dd X^-\,.
\end{align}

\medskip 

For the background dilaton $\bar{\phi}$\,, it is helpful to take a particular choice 
of $a$, $b$ and $c$ as  
\begin{align}
a=-\frac{\Lambda\,L}{\sqrt{2}}\,,\qquad b=0\,,\qquad c=\frac{\sqrt{2}\Lambda}{L}\,. 
\label{choice}
\end{align}
Then the dilaton is rewritten as 
\begin{align}
\bphi=\frac{\Lambda\,L^2}{2}
\left(1-\frac{1}{\sqrt{2}\,L}\frac{L^2-2\,x^+ x^-}{x^+-x^-}\right)\,.
\end{align}
After taking the limit $L\rightarrow\infty$\,, the dilaton reduces to the one (\ref{eq:flat-bg}) 
in the flat-space JT gravity: 
\begin{align}
\bphi \quad \longrightarrow \quad \frac{\Lambda}{2}X^+X^-\,.
\end{align}

\medskip 

It may be worth noting that the choice (\ref{choice}) corresponds to a black hole solution 
discussed in \cite{AP}. In particular, the parameter $c$ is basically associated with 
the Hawking temperature and eventually this part gives rise to the dilaton in the flat-space JT gravity. 
It would be intriguing to try to get much deeper understanding for this connection.

\subsubsection*{(ii) 2D de Sitter space}

An another interesting example is  a de Sitter version of the AP model. 
This case is realized by taking the following dilaton potential, 
\begin{align}
U(\phi)=\Lambda+\frac{2}{L^2}\phi\,.
\end{align}
For the recent progress on the dS$_2$ in the JT gravity, see \cite{dS2,Jensen2}. 
There might be a potential application in the context of dS/dS correspondence \cite{Silverstein}. 

\medskip 

The general vacuum solution incorporates the dS$_2$ metric and a non-constant dilaton
\begin{align}
{\rm e}^{2\bar{\omega}}
= \frac{2\,L^2}{(x^++x^-)^2}\,,\qquad \bphi=\frac{L^2}{2}
\left(-\Lambda+\frac{a+b\,(x^+ -x^-)+c\,x^+ x^-}{x^++x^-}\right)\,, 
\end{align}
where $L$ is the curvature radius and $a$, $b$ and $c$ are arbitrary constants.

\medskip 

Again, let us examine the equations of motion  (\ref{eq:eom-usefull}). 
The $(++)$ and $(--)$ components of (\ref{eq:eom-usefull}) are evaluated as 
\begin{align}
{\rm e}^{2\bar{\omega}}\partial_+\left({\rm e}^{-2\bar{\omega}}\partial_+\sigma\right)=
&  -8\pi G_N {{\cal T}_{\rm dS}}_+(x^+)\,,
\label{eq:dS++}\\
{\rm e}^{2\bar{\omega}}\partial_-\left({\rm e}^{-2\bar{\omega}}\partial_-\sigma\right)=&
 -8\pi G_N {{\cal T}_{\rm dS}}_-(x^-)\,, 
\label{eq:dS--}
\end{align}
where ${{\cal T}_{\rm dS}}_{\pm} (x^{\pm})$ are defined as 
\begin{equation}
{{\cal T}_{\rm dS}}_{\pm} (x^{\pm}) \equiv(1 \pm b\,k - c\,k\,x^{\pm})\,t_{\pm \pm} 
+ \frac{k}{4}\left(a \pm2b\,x^{\pm} - c\,(x^{\pm})^2\right)\partial_{\pm} t_{\pm \pm}\,. 
 \end{equation}
Similarly to the AdS case, it is useful to represent $\sigma$ by using a scalar function 
$M_{\rm dS}(x^+\,,x^-)$:
\begin{align}
\sigma(x^+,x^-) \equiv \frac{M_{\rm dS}(x^+\,,x^-)}{x^++x^-}\,.
\end{align}
By integrating (\ref{eq:dS++}) and (\ref{eq:dS--}), the general solution can be derived as 
\begin{align}
M_{\rm dS}(x^+\,,x^-)=J_0(x^+,x^-) + J^+(x^+,x^-) + J^-(x^+,x^-)\,.\label{dSsolution}
\end{align}
Here $J_0$ is the sourceless solution 
\[
J_0(x^+,x^-)\equiv A + B\,(x^+-x^-) + C\, x^+x^-\,,  \qquad A,~B,~C \mbox{:~arbitrary real consts.}, 
\] 
and $J^\pm(x^+,x^-)$ correspond to the non-local part of dilaton and are given by 
\begin{align}
J^\pm(x^+,x^-) \equiv &\, 8\pi G_N \int^{x^\pm}_{u^\pm}\!\!\dd s\, 
(s\mp x^+)(s\pm x^-)\,{{\cal T}_{\rm dS}}_\pm(s)\,.
\end{align}
The $(+-)$ component of (\ref{eq:eom-usefull}) is drastically simplified 
due to the traceless condition $t_{+-}=0$ and is given by 
\begin{align}
\partial_+\partial_-\sigma-\frac{2\,\sigma}{(x^++x^-)^2}=&\,0\,.\label{eq:dS+-}
\end{align}
$\sigma$ with (\ref{dSsolution}) also satisfies the above condition (\ref{eq:dS+-}).

\medskip

\subsubsection*{A flat-space limit}

In a similar way to the AdS$_2$ case, it is easy to consider a flat-space limit 
for the vacuum solution in the dS model. 

\medskip

Let us take a constant shift of $x^\pm$ and introduce 
new coordinates $X^\pm$ defined as 
\begin{align}
X^\pm \equiv x^\pm \pm\frac{L}{\sqrt{2}}\,.
\end{align}
By taking the large radius limit $L\rightarrow\infty$, the dS$_2$ metric goes 
to the Minkowski metric, 
\begin{align}
{\rm d}s^2
= \frac{-4\,L^2\dd x^+\dd x^-}{(x^++x^-)^2} \quad 
\longrightarrow \quad 
-2\,\dd X^+\dd X^-\,.
\end{align}

\medskip 

For the background dilaton $\bar{\phi}$\,, it is helpful to take a particular choice 
of $a$, $b$ and $c$ as  
\begin{align}
a=\frac{\Lambda\,L}{\sqrt{2}}\,,\qquad b=0\,,\qquad c=\frac{\sqrt{2}\Lambda}{L}\,. 
\end{align}
Then the dilaton is rewritten as 
\begin{align}
\bphi=\frac{\Lambda\,L^2}{2}
\left(-1+\frac{1}{\sqrt{2}\,L}\frac{L^2+2\,x^+ x^-}{x^++x^-}\right)\,.
\end{align}
After taking the limit $L\rightarrow\infty$\,, the dilaton reduces to the one (\ref{eq:flat-bg}) 
in the flat-space JT gravity: 
\begin{align}
\bphi \quad \longrightarrow \quad \frac{\Lambda}{2}X^+X^-\,.
\end{align}

\section{Conclusion and discussion}

In this paper, we have discussed gravitational perturbations in 2D dilaton-gravity systems 
with some dilaton potentials, motivated by the pioneering work \cite{Dubovsky}. 
It has been shown that at least under some conditions the perturbations can be regarded 
as $T\bar{T}$-deformations of the original matter action. In particular, the class of theories 
that this identification applies to includes the JT gravity with a cosmological constant,
and there would be potential applications in the context of the AdS$_2$ or dS$_2$ holography. 
It is significant to figure out to what extent this identification should be valid, but we will leave it 
as a future problem. 

\medskip 

There are some open problems and future directions. Here we have discussed the case of 
the gravitational perturbation in the case of the JT gravity including a negative cosmological constant. 
It should be important to study the AdS$_2$ holography from the viewpoint of 
the $T\bar{T}$-deformation of the bulk geometry. 
Along this direction, the Gibbons-Hawking term, which has not been 
taken into account here, should be considered carefully so that one can seek for  
the CFT$_1$ interpretation of this gravitational $T\bar{T}$-deformation. In addition, 
it will also be important to compute the partition function by following \cite{quantum}. 
In this evaluation, the spacetime has to be compactified so as to obtain the finite-volume spectrum. 
While in the flat-space JT gravity the spacetime is compactified to a torus (with the Wick rotation), 
what should we do for the AdS$_2$ case? We anticipate that the answer would be to utilize 
the cut-off AdS geometry \cite{Verlinde} and we hope to report on some progress 
along this direction in the near future.  

\medskip 

The most significant issue is to investigate gravitationally dressed S-matrix in the matter field theory. 
In the case of the flat-space JT gravity, a dressing factor to the S-matrix caused 
by the $T\bar{T}$-deformation has been evaluated \cite{Dubovsky}. At least in principle, it should be 
possible to carry out similar analysis and examine the dressing factor, though the treatment of 
the S-matrix on a curved spacetime would be quite complicated. As a matter of course, 
the 2D de Sitter case would be intriguing. It is also interesting to consider a relation to 
the random geometry by following the work \cite{Cardy}.

\medskip 

It is also interesting to try to relax our condition so as to support non-conformal matter fields,  
more general types of dilaton potential, and general forms of dilaton coupling to matter fields. 
In particular, it has been shown that Yang-Baxter deformations \cite{Klimcik, DMV,KMY} 
are closely related to $T\bar{T}$-deformations \cite{Araujo,Borsato} (For related works, 
see \cite{Sfondrini1,Sfondrini2} as well). Hence Yang-Baxter deformations 
of the JT gravity \cite{KOY,regular,Frolov} may also be related to $T\bar{T}$-deformations. 
It would also be nice to generalize the relation between the deformed solutions and 
the unperturbed ones via a specific field-dependent local change of coordinates, 
which has been revealed in \cite{Tateo2,Tateo3}, from flat space to curved spaces 
like AdS$_2$ or dS$_2$. 
As another direction, it may be interesting to try to include supersymmetries 
by following the works \cite{SUSY1,SUSY2,SUSY3}.

\medskip 

We hope that our result would open up a new route to the AdS$_2$ holography.

\subsection*{Acknowledgments}

We thank H.~Y.~Chen, D.~Orlando, S.~Reffert and R.~Tateo for useful discussions.  
In particular, K.Y.\ is really grateful to R.~Tateo for drawing our attention to this subject 
and also would like to thank discussions with participants during the OIST workshop on 
``Recent developments in AdS/CFT,'' organized by N.~Iizuka and T.~Ugajin. 
The work of T.I.\ was supported in part by JSPS Grant-in-Aid for Scientific Research (C) No.\,19K03871.
The work of S.O.\ was supported by the Japan Society for the Promotion of Science (JSPS). 
The work of J.S.\ was supported in part by JSPS Grant-in-Aid for Scientific Research (B) No.\,16H03979 and Osaka City University Advanced Mathematical Institute (MEXT Joint Usage/Research Center on Mathematics and Theoretical Physics).
The works of T.I.\ and K.Y.\ were supported by the Supporting Program for Interaction-based Initiative 
Team Studies (SPIRITS) from Kyoto University, and JSPS Grant-in-Aid for Scientific Research (B) 
No.\,18H01214.
This work is also supported in part by the JSPS Japan-Russia Research Cooperative Program.

\appendix

\section*{Appendix}

\section{Useful formulae}\label{sec:Ricci}

It would be helpful for readers to summarize formulae useful in computing some quantities in this paper. 

\medskip 

We consider a small perturbation around a given metric as 
\begin{align}
g_{\mu\nu}=\bar{g}_{\mu\nu}+h_{\mu\nu}\,,
\end{align}
and expand some geometric quantities in terms of the metric fluctuation 
up to and including the second-order in $h_{\mu\nu}$. 
In the following we will drop the higher order terms. 
The explicit expressions of the perturbed quantities are useful in deriving the quadratic action 
(\ref{eq:quadratic-action}).

\medskip

We start by expanding the inverse and the determinant of the perturbed metric $g_{\mu\nu}$\,.
These are given by
\begin{align}
g^{\mu\nu}& = \bar{g}^{\mu\nu}-h^{\mu\nu}+h^{\mu}_{\rho}h^{\rho\nu}\,,\label{eq:invg}\\
\sqrt{-g}&=\sqrt{-\bar{g}}\left(1+\frac{1}{2}h-\frac{1}{4}\left(h_{\mu\nu}h^{\mu\nu}-\frac{1}{2}h^2\right)\right)\,,
\label{eq:Detg}
\end{align}
where $h^{\mu\nu}=\bar{g}^{\mu\rho}\bar{g}^{\nu \sigma}h_{\rho\sigma}$\,, $h=\bar{g}^{\mu\nu}h_{\mu\nu}=h^{\mu}_{\mu}$\,. 

\medskip

The Christoffel symbol is defined as 
\begin{align}
\Gamma^{\rho}_{\mu\nu}\equiv
\frac{1}{2}g^{\rho\sigma}
(\partial_{\mu} g_{\sigma\nu}+\partial_{\nu} g_{\sigma\mu}-\partial_{\sigma} g_{\mu\nu})\,,
\end{align}
and can be expanded as
\begin{align}
\Gamma^{\rho}_{\mu\nu}=
\bar{\Gamma}^{\rho}_{\mu\nu} + \Gamma^{(1)\rho}_{\mu\nu} + \Gamma^{(2)\rho}_{\mu\nu}\,,
\end{align}
where the first and second order terms in the fluctuation are
\begin{align}
\Gamma^{(1)\rho}_{\mu\nu}&=
\frac{1}{2} \bar{g}^{\rho\sigma}
(\bar{\nabla}_{\mu} h_{\sigma\nu} + \bar{\nabla}_{\nu} h_{\sigma\mu} 
- \bar{\nabla}_{\sigma} h_{\mu\nu})\,,\\
\Gamma^{(2)\rho}_{\mu\nu}&=
-\frac{1}{2}h^{\rho\sigma}
(\bar{\nabla}_{\mu} h_{\sigma\nu}+\bar{\nabla}_{\nu} h_{\sigma\mu}-\bar{\nabla}_{\sigma} h_{\mu\nu})
\,.
\end{align}

\medskip

The Riemann tensor and the Ricci tensor are defined as
\begin{align}
R^{\mu}{}_{\nu\alpha\beta}&
\equiv\partial_{\alpha} \Gamma^{\mu}_{\nu\beta}-\partial_{\beta}\Gamma^{\mu}_{\nu\alpha}
+\Gamma^{\mu}_{\rho\alpha}\Gamma^{\rho}_{\nu\beta}
-\Gamma^{\mu}_{\rho\beta}\Gamma^{\rho}_{\nu\alpha}\,,\\
R_{\mu\nu}&\equiv R^{\rho}{}_{\mu\rho\nu}\,.
\end{align}
The Ricci tensor can be expanded as 
\begin{align}
R_{\mu\nu} = \bar{R}_{\mu\nu} + R^{(1)}_{\mu\nu} + R^{(2)}_{\mu\nu}\,,
\end{align}
where $R^{(1)}_{\mu\nu}$ and $R^{(2)}_{\mu\nu}$ are 
\begin{align}
R^{(1)}_{\mu\nu}&=\frac{1}{2}\bar{\nabla}^{\rho}\left(\bar{\nabla}_{\mu}h_{\rho \nu}+\bar{\nabla}_{\nu} h_{\rho \mu}-\bar{\nabla}_{\rho}h_{\mu\nu}\right)
-\frac{1}{2}\bar{\nabla}_{\mu}\bar{\nabla}_{\nu} h\,,\label{eq:Riccit-first}\\
R^{(2)}_{\mu\nu}&=\frac{1}{2}\bar{\bar{\nabla}}_{\nu}(h^{\rho\sigma}\bar{\bar{\nabla}}_{\mu}h_{\rho\sigma})-\frac{1}{2}\bar{\bar{\nabla}}_{\rho}\left[h^{\rho\sigma}
\left(\bar{\nabla}_{\mu}h_{\sigma\nu}+\bar{\nabla}_{\nu} h_{\sigma \mu}-\bar{\nabla}_{\sigma}h_{\mu\nu}\right)\right]\no\\
&\quad+\frac{1}{4}\bar{\nabla}^\rho h\left(\bar{\nabla}_{\mu}h_{\rho \nu}+\bar{\nabla}_{\nu} h_{\rho \mu}-\bar{\nabla}_{\rho}h_{\mu\nu}\right)\no\\
&\quad-\frac{1}{4} \bar{g}^{\alpha\beta} \bar{g}^{\rho\sigma}
\left(\bar{\nabla}_{\mu}h_{\alpha \rho}+\bar{\nabla}_{\alpha} h_{\rho \mu}-\bar{\nabla}_{\rho}h_{\alpha\mu}\right)
\left(\bar{\nabla}_{\sigma}h_{\beta \nu}+\bar{\nabla}_{\nu} h_{\beta \sigma}-\bar{\nabla}_{\beta}h_{\sigma\nu}\right)\,.
\label{eq:Riccit-second}
\end{align}

\medskip

The Ricci scalar $R=g^{\mu\nu}R_{\mu\nu}$ can be expanded as
\begin{align}
R = \bar{R}+R^{(1)}+R^{(2)}\,,
\end{align}
where $R^{(1)}$ and $R^{(2)}$ are given by
\begin{align}
R^{(1)}& = \bar{g}^{\mu\nu}R_{\mu\nu}^{(1)}-h^{\mu\nu} \bar{R}_{\mu\nu}\no\\
&=\bar{\nabla}^{\mu}\bar{\nabla}^{\nu}h_{\mu\nu}-\bar{\nabla}^2 h-h^{\mu\nu} \bar{R}_{\mu\nu}\,,\label{eq:Riccis-first}\\
R^{(2)}& = \bar{g}^{\mu\nu} R^{(2)}_{\mu\nu}-h^{\mu\nu}R^{(1)}_{\mu\nu}
+h^{\mu}_{\rho}h^{\rho\nu} \bar{R}_{\mu\nu}\,,\label{eq:Riccis-quadratic}
\end{align}
with
\begin{align}
\bar{g}^{\mu\nu} R^{(2)}_{\mu\nu}&=\frac{1}{2}\bar{\nabla}^{\mu}(h^{\rho\sigma}\bar{\nabla}_{\mu}h_{\rho\sigma})
-\bar{\nabla}_{\rho}\left[h^{\rho\sigma}
\left(\bar{\nabla}^{\mu}h_{\sigma\mu}-\frac{1}{2}\bar{\nabla}_{\sigma}h\right)\right]\no\\
&\quad+\frac{1}{2}\bar{\nabla}^\rho h\left(\bar{\nabla}^{\mu}h_{\rho\mu}-\frac{1}{2}\bar{\nabla}_{\rho}h\right)\no\\
&\quad-\frac{1}{4} \bar{g}^{\mu\nu} \bar{g}^{\alpha\beta} \bar{g}^{\rho\sigma}
\left(\bar{\nabla}_{\mu}h_{\alpha \rho}+\bar{\nabla}_{\alpha} h_{\rho \mu}-\bar{\nabla}_{\rho}h_{\alpha\mu}\right)
\left(\bar{\nabla}_{\sigma}h_{\beta \nu}+\bar{\nabla}_{\nu} h_{\beta \sigma}-\bar{\nabla}_{\beta}h_{\sigma\nu}\right)\no\\
&\quad-h^{\mu\nu}R^{(1)}_{\mu\nu}
+h^{\mu}_{\rho}h^{\rho\nu} \bar{R}_{\mu\nu}
\no\\
&=\bar{\nabla}^{\rho}\left(\frac{3}{4}h^{\mu\nu}\bar{\nabla}_{\rho}h_{\mu\nu}
-h_{\rho}^{\sigma}
\left(\bar{\nabla}^{\mu}h_{\sigma\mu}-\frac{1}{2}\bar{\nabla}_{\sigma}h\right)\right)
+\frac{1}{2}\bar{\nabla}^\rho h\left(\bar{\nabla}^{\mu}h_{\rho\mu}-\frac{1}{2}\bar{\nabla}_{\rho}h\right)\no\\
&\quad-\frac{1}{2}\bar{\nabla}_{\mu}h_{\alpha\beta} \bar{\nabla}^{\alpha}h^{\beta\mu}
-\frac{1}{4}h^{\mu\nu} \bar{\nabla}^2h_{\mu\nu}
-h^{\mu\nu}R^{(1)}_{\mu\nu}
+h^{\mu}_{\rho}h^{\rho\nu} \bar{R}_{\mu\nu}\,.\label{eq:Riccis-quadratic2}
\end{align}

\section{A derivation of the quadratic action}\label{sec:der-qaction}

We explain how to derive the quadratic action (\ref{eq:quadratic-action}) in detail. 

\medskip

By using (\ref{eq:invg}) and (\ref{eq:Detg}), it is straightforward to derive the following quadratic action:
\begin{align}
S^{(2)}&=\frac{1}{16\pi G_N}\int\! \dd^2 x\,\sqrt{-\bar{g}}\,\biggl(\left[\frac{1}{2}h\,\left(\bar{R}-U'(\bar{\phi})\right)+R^{(1)}-\frac{1}{2}U''(\bar{\phi}) \sigma\right]\sigma\no\\ 
&\qquad+\frac{1}{4}U(\bar{\phi})\left(h_{\mu\nu}h^{\mu\nu}-\frac{1}{2}h^2\right)
+\bar{\phi}\left[R^{(2)}+\frac{1}{2} h R^{(1)}-\frac{1}{4}\bar{R}\left(h_{\mu\nu}h^{\mu\nu}-\frac{1}{2}h^2\right)\right]\biggr)\no\\
&\qquad+\frac{1}{2}\int\! \dd^2 x\,\sqrt{-\bar{g}}\,h^{\mu\nu} t_{\mu\nu}\,.
\label{eq:action-second-1}
\end{align}
By using the explicit expression (\ref{eq:Riccis-first}) of $R^{(1)}$ and the vanishing of the two dimensional Einstein tensor (\ref{eq:2dEin}), this can be rewritten as
\begin{align}
S^{(2)}&=\frac{1}{16\pi G_N}\int\! \dd^2 x\,\sqrt{-\bar{g}}\,
\biggl(\left[\bar{\nabla}^{\mu}\bar{\nabla}^{\nu}h_{\mu\nu}-\bar{\nabla}^2 h-\frac{1}{2}h\,U'(\bar{\phi})
-\frac{1}{2}U''(\bar{\phi}) \sigma\right]\sigma\no\\
&\qquad+\frac{1}{4}U(\bar{\phi})\left(h_{\mu\nu}h^{\mu\nu}-\frac{1}{2}h^2\right)
+\bar{\phi}\left[R^{(2)}+\frac{1}{2} h R^{(1)}-\frac{1}{4} \bar{R}\left(h_{\mu\nu}h^{\mu\nu}-\frac{1}{2}h^2\right)\right]\biggr)\no\\
&\qquad+\frac{1}{2}\int\! \dd^2 x\,\sqrt{-\bar{g}}\,h^{\mu\nu} t_{\mu\nu}\,.
\label{eq:action-second-2}
\end{align}

We then rewrite the terms proportional to the background dilaton $\bar{\phi}$ 
in (\ref{eq:action-second-2}).
For this purpose, it is helpful to employ the following identity:
\begin{align}
0=h^{\mu\nu}\left(R_{\mu\nu}-\frac{1}{2}g_{\mu\nu}R\right)\,.
\label{eq:hEin}
\end{align}
The identity (\ref{eq:hEin}) gives at $\mathcal{O}(h_{\mu\nu}^2)$
\begin{align}
-\frac{1}{4}\bar{R}\left(h_{\mu\nu}h^{\mu\nu}-\frac{1}{2}h^2\right)
=-\frac{1}{2}h^{\mu\nu}R^{(1)}_{\mu\nu}+\frac{1}{4}h\left(\bar{\nabla}^{\mu}\bar{\nabla}^{\nu}h_{\mu\nu}-\bar{\nabla}^{2}h\right)\,.
\label{eq:Einten-2h}
\end{align}
By using the identity (\ref{eq:Einten-2h}) and the formula (\ref{eq:Riccis-quadratic}) together with 
the fact that the Einstein tensor vanishes in two dimensions (\ref{eq:2dEin}), 
the part proportional to the background dilaton $\bar{\phi}$ in (\ref{eq:action-second-2}) 
can be rewritten as
\begin{align}
&\bar{\phi}\left[R^{(2)}+\frac{1}{2} h R^{(1)}-\frac{1}{4}\bar{R}\left(h_{\mu\nu}h^{\mu\nu}-\frac{1}{2}h^2\right)\right]\no\\
&=\bar{\phi}\left[\bar{g}^{\mu\nu}R^{(2)}_{\mu\nu}-\frac{1}{2}h^{\mu\nu}R^{(1)}_{\mu\nu}
+\frac{1}{4}h\left(\bar{\nabla}^{\mu}\bar{\nabla}^{\nu}h_{\mu\nu}-\bar{\nabla}^2 h\right)\right]\,.
\label{eq:phi-pro1}
\end{align}
Substituting (\ref{eq:Riccit-first}) and (\ref{eq:Riccis-quadratic2}) into (\ref{eq:phi-pro1}),
the above expression becomes
\begin{align}
({\rm \ref{eq:phi-pro1}})&=\bar{\phi}\bar{\nabla}^\rho\biggl(\frac{3}{4}h^{\mu\nu}\bar{\nabla}_{\rho}
h_{\mu\nu}
-\frac{1}{4}h\bar{\nabla}_{\rho}h
-h_{\rho\sigma} \bar{\nabla}_{\mu}h^{\mu\sigma}\no\\
&\qquad\qquad-\frac{1}{2}h^{\mu\nu}\bar{\nabla}_{\mu}h_{\rho \nu}+\frac{1}{4}h \bar{\nabla}^{\sigma}h_{\sigma\rho}
+\frac{3}{4}h_{\rho\mu}\bar{\nabla}^{\mu}h
\biggr)\no\\
&=\frac{1}{4}\bar{\nabla}^{2}\bar{\phi}\left(h_{\mu\nu}h^{\mu\nu}-\frac{1}{2}h^2\right)
+\frac{1}{8}\bar{\nabla}^{2}\bar{\phi}\,h_{\mu\nu}h^{\mu\nu}\no\\
&\quad + \bar{\phi} \bar{\nabla}^\rho\biggl(
-h_{\rho\sigma} \bar{\nabla}_{\mu}h^{\mu\sigma}
-\frac{1}{2}h^{\mu\nu}\bar{\nabla}_{\mu}h_{\rho \nu}+\frac{1}{4}h \bar{\nabla}^{\sigma}h_{\sigma\rho}
+\frac{3}{4}h_{\rho\mu}\bar{\nabla}^{\mu}h
\biggr)\no\\
&\quad + \bar{\nabla}^{\rho}\left(\bar{\phi}\left[\frac{3}{4}h^{\mu\nu}\bar{\nabla}_{\rho}h_{\mu\nu}
-\frac{1}{4}h \bar{\nabla}_{\rho}h\right]
-\bar{\nabla}_{\rho}\bar{\phi}\left[\frac{3}{8}h^{\mu\nu}-\frac{1}{8}h^2\right]\right)
\,.
\label{eq:phi-pro2}
\end{align}
Furthermore, it should be useful to rewrite the second line from the end in (\ref{eq:phi-pro2})
so that it contains only the terms containing $\bar{\nabla}_{\mu}h$ and $\bar{\nabla}^{\mu}h_{\mu\nu}$\,.
By using the fact that the background dilaton $\bar{\phi}$ satisfies (\ref{eq:eom-phimunu}),
the second term in that line can be rewritten as 
\begin{align}
-\frac{1}{2}\bar{\phi}  
\bar{\nabla}^{\rho}(h^{\mu\nu}\bar{\nabla}_{\mu}h_{\rho\nu})
&=-\frac{1}{2}(\bar{\nabla}_{\mu}\bar{\nabla}^{\rho}\bar{\phi})h^{\mu\nu}h_{\rho\nu}
-\frac{1}{2}\bar{\nabla}^{\rho}\bar{\phi} \,h_{\rho\nu}\bar{\nabla}_{\mu}h^{\mu\nu}\no\\
&\quad-\frac{1}{2}\bar{\nabla}^{\rho}(\bar{\phi}\,h^{\mu\nu}\bar{\nabla}_{\mu}h_{\rho\nu})
+\frac{1}{2}\bar{\nabla}_{\mu}(\bar{\nabla}^{\rho}\bar{\phi}\,h^{\mu\nu} h_{\rho\nu})
\no\\
&
=-\frac{1}{4}\bar{\nabla}^2 \bar{\phi}\,h_{\mu\nu}h^{\mu\nu}
-\frac{1}{2}\bar{\nabla}^{\rho} \bar{\phi} \,
h_{\rho\nu} \bar{\nabla}_{\mu}h^{\mu\nu}
\no\\
&\quad-\frac{1}{2}\bar{\nabla}^{\rho}(\bar{\phi}\,h^{\mu\nu}\bar{\nabla}_{\mu}h_{\rho\nu})
+\frac{1}{2}\bar{\nabla}_{\mu}(\bar{\nabla}^{\rho}\bar{\phi}\,h^{\mu\nu} h_{\rho\nu})
\,.
\end{align}
As a result, we obtain
\begin{align}
&\bar{\phi}\left[R^{(2)}+\frac{1}{2} h R^{(1)}-\frac{1}{4}\bar{R}\left(h_{\mu\nu}h^{\mu\nu}-\frac{1}{2}h^2\right)\right]\no\\
&=\frac{1}{4}\bar{\nabla}^{2}\bar{\phi}\left(h_{\mu\nu}h^{\mu\nu}-\frac{1}{2}h^2\right)
-\frac{1}{8}\bar{\nabla}^{2}\bar{\phi}\,h_{\mu\nu}h^{\mu\nu}
-\frac{1}{2}\bar{\nabla}^{\rho}\bar{\phi} \,
h_{\nu\rho}\bar{\nabla}_{\mu}h^{\mu\nu}\no\\
&\quad+\bar{\phi}\bar{\nabla}^\rho\left(
-h_{\rho\sigma} \bar{\nabla}_{\mu}h^{\mu\sigma}+\frac{1}{4}h \bar{\nabla}^{\sigma}h_{\sigma\rho}
+\frac{3}{4}h_{\rho\mu}\bar{\nabla}^{\mu}h
\right)
-\frac{1}{2}\bar{\nabla}^{\rho}\bar{\phi} \,
h_{\rho\nu}\bar{\nabla}_{\mu}h^{\mu\nu}
\no\\
&\quad+\bar{\nabla}^{\rho}\left(\bar{\phi}\left[\frac{3}{4}h^{\mu\nu}\bar{\nabla}_{\rho}h_{\mu\nu}
-\frac{1}{4}h \bar{\nabla}_{\rho}h
-\frac{1}{2}h^{\mu\nu} \bar{\nabla}_{\mu}h_{\nu\rho}\right]
-\bar{\nabla}_{\rho}\bar{\phi}\left[\frac{3}{8}h^{\mu\nu}-\frac{1}{8}h^2\right]
\right)\no\\
&\quad+\frac{1}{2}\bar{\nabla}_{\mu}(\bar{\nabla}^{\rho}\bar{\phi}\,h^{\mu\nu} h_{\nu\rho})
\,.\label{eq:Einten-2h-2}
\end{align}

\medskip

Finally, by using (\ref{eq:Einten-2h-2}) and the on-shell condition of the background dilaton 
(\ref{eq:eom-phimunu}) and doing partial integration, 
the quadratic action $S^{(2)}$ becomes
\begin{align}
S^{(2)}&=\frac{1}{16\pi G_N}\int\! \dd^2 x\,
\sqrt{-\bar{g}}\,\biggl(\left[\bar{\nabla}^{\mu}\bar{\nabla}^{\nu}h_{\mu\nu}-\bar{\nabla}^2 h 
-\frac{1}{2}h\,U'(\bar{\phi})
-\frac{1}{2}U''(\bar{\phi}) \sigma\right]\sigma
\no\\
&\qquad
-\frac{1}{8}\bar{\nabla}^{2}\bar{\phi}\,h_{\mu\nu}h^{\mu\nu}
-\bar{\nabla}^{\rho}\bar{\phi} \left[
-\frac{1}{2}h_{\rho\sigma} \bar{\nabla}_{\mu}h^{\mu\sigma}
+\frac{1}{4}h \bar{\nabla}^{\sigma}h_{\sigma\rho}
+\frac{3}{4}h_{\rho\mu}\bar{\nabla}^{\mu}h
\right]\biggr)\no\\
&\qquad+\frac{1}{2}\int\! \dd^2 x\,\sqrt{-\bar{g}}\,h^{\mu\nu} t_{\mu\nu}\,,
\end{align}
where we have ignored the total derivative terms.
This is the quadratic action in (\ref{eq:quadratic-action}).

\end{document}